\documentclass[aps,prd,preprint,superscriptaddress]{revtex4-1}
\usepackage[mathscr]{euscript}
\usepackage{float,epsfig}
\usepackage{color,soul}
\usepackage[normalem]{ulem}
\usepackage{bm}% bold math
\usepackage{graphicx}% Include figure files
\usepackage{subfigure}
\usepackage{appendix}
\usepackage{amsmath,amssymb,amsthm}
\usepackage[colorlinks=true,linkcolor=blue,urlcolor=blue]{hyperref}
\newcommand{\bea}{\begin{eqnarray}}
\newcommand{\eea}{\end{eqnarray}}
\newcommand{\beq}{\begin{equation}}
\newcommand{\eeq}{\end{equation}}

\def\/{\over}
\allowdisplaybreaks[4]

\begin{document}

%\preprint{}

\title{Quantum gravitomagnetic interaction}

\author{Di Hao}
\affiliation{Department of Physics, Synergetic Innovation Center for Quantum Effects and Applications,\\ and Institute of Interdisciplinary Studies, \\Hunan Normal University, Changsha, Hunan 410081, China}
\author{Jiawei Hu}
\email[Corresponding author. ]{jwhu@hunnu.edu.cn}
\affiliation{Department of Physics, Synergetic Innovation Center for Quantum Effects and Applications,\\ and Institute of Interdisciplinary Studies, \\Hunan Normal University, Changsha, Hunan 410081, China}
\author{Hongwei Yu}
\email[Corresponding author. ]{hwyu@hunnu.edu.cn}
\affiliation{Department of Physics, Synergetic Innovation Center for Quantum Effects and Applications,\\ and Institute of Interdisciplinary Studies, \\Hunan Normal University, Changsha, Hunan 410081, China}

\begin{abstract}

In the framework of linearized quantum gravity, we study the quantum gravitational interaction between two nonpointlike objects induced by fluctuating gravitomagnetic fields in vacuum. We find that, in addition to the quantum gravitational interaction induced by fluctuating gravitoelectric fields previously studied, there exists a quantum gravitomagnetic interaction. This interaction originates from the interaction between the instantaneous localized mass currents in nonpointlike objects induced by the fluctuating gravitomagnetic fields. Using fourth-order perturbation theory, we derive the explicit form of the quantum gravitomagnetic interaction energy, which shows an $r^{-10}$ dependence in the near regime and an $r^{-11}$ dependence in the far regime, where $r$ is the distance between the two objects. This interaction energy is expected to be significant when the gravitomagnetic polarizability of the objects is large.

\end{abstract}

%\pacs{}
%\keywords{}

\maketitle

\section{Introduction}
\label{sec_in}
\setcounter{equation}{0}
%%%%%%%%%%%%%%%%%%%%%%%%%%%%%%%%%%%%%%%%%%%%%%%%%%
Although a theory of quantum gravity, which is necessary to understand quantum gravitational effects near the Planck energy scale,  is  elusive, one can still study quantum gravitational effects  at low-energy scales, for an example, by taking general relativity as an effective field theory. In such an effective field approach, it has been shown that  there exists a quantum correction to the classical Newtonian potential, which can be obtained by summing one-loop Feynman diagrams involving off-shell gravitons~\cite{Donoghue1994prl,Donoghue1994prd,Hamber1995,Kirilin2002,Holstein2003,Holstein2005},  just as the consideration of the radiative corrections of quantum electrodynamics would lead to a modification of the Coulomb interaction between two charges (See, e.g., Ref. \cite{Bjorken64}).

The works mentioned above deal with the quantum gravitational interaction between two mass monopoles. Recently, the quantum gravitational interaction between two nonpointlike objects has also been investigated~\cite{Wu2016,Ford2016,Holstein2017}. This interaction originates from the interaction between instantaneous mass quadrupoles induced by the fluctuating gravitational fields, and behaves as $r^{-10}$ and $r^{-11}$ in the near and far regimes, respectively. Due to its close analogy to the Casimir-Polder interaction between a pair of atoms due to the interaction between instantaneous electric dipoles induced by the fluctuating electromagnetic fields~\cite{Casimir1948pr}, it is also dubbed as the gravitational Casimir-Polder interaction. Originally, this quantum gravitational interaction is derived with a method in close analogy to that in the computation of the electromagnetic Casimir-Polder interaction between two atoms from their induced  electric dipole moments due to two-photon exchange~\cite{Ford2016}, in which the details of quantization of the gravitational field are not needed. The result has soon been confirmed by the leading-order perturbation calculations based on linearized quantum gravity~\cite{Wu2016}, and the calculation of scattering amplitudes~\cite{Holstein2017}. Later, in the framework of linearized quantum gravity,  the quantum gravitational interaction on a nonpointlike object  and between two nonpointlike objects near a boundary~\cite{Hu2017,yu2018}, as well as the interaction between two nonpointlike objects in a thermal bath \cite{Wu2017},  a pair of objects in the symmetric or antisymmetric entangled state~\cite{yongs2020epjc}, and among three nonpointlike objects~\cite{yongs2022prd}  has also been studied.

Under weak-field approximation, the linearized Einstein field equations can be organized in a form similar to the Maxwell equations, in which the gravitoelectric field and the gravitomagnetic field play the roles similar to those played by the electric and magnetic fields in electromagnetism respectively. This is known as Weyl gravitoelectromagnetism~\cite{Campbell1976,Matte1953,Campbell1971,Szekeres1971,Maartens1998,Ruggiero2002,Ramos2010}. In the framework of the gravitoelectromagnetism, the instantaneous mass quadrupole moments responsible for the gravitational Casimir-Polder interaction are induced by the fluctuating gravitoelectric fields~\cite{Wu2016,Wu2017,Ford2016,Holstein2017,yu2018,Hu2017,yongs2020epjc,yongs2022prd}. 
A question then naturally arises as to what happens if the fluctuating gravitomagnetic fields are considered. 
Here, let us note that the Casimir-Polder interaction between a pair of neutral but polarizable atoms due to the interaction between instantaneous magnetic dipoles induced by the fluctuating electromagnetic fields (in natural units $c=\hbar=1$) is of the same form as that due to the interaction between instantaneous electric dipoles, apart from the replacement of electric polarizability with magnetic susceptibility~\cite{Buhmann2005,Buhmann_book,Salam2010}. Therefore, we are particularly concerned about whether such a correspondence still exists in the gravitational case. 
This is what we are going to investigate in the present paper. The paper is organized as follows. In Sec. \ref{sec_ge}, we derive the interaction Hamiltonian describing the interaction between an object and the fluctuating gravitomagnetic fields. In Sec. \ref{sec3}, we calculate the quantum gravitomagnetic interaction between two nonpointlike objects  induced by the fluctuating gravitomagnetic fields using the fourth-order perturbation theory, and analyze its asymptotic behaviors in the near and far regimes. We summarize in Sec. \ref{sec_disc}. Throughout the paper, the Greek indices take values from 0 to 3, and the Latin indices run from 1 to 3. The Einstein summation convention is assumed for repeated indices. Unless otherwise specified, the natural units $c=\hbar=1$ are adopted.

%%%%%%%%%%%%%%%%%%%%%%%%%%%%%%%%%%%%%%%%%%%%%%%%%%
\section{The interaction Hamiltonian}
\label{sec_ge}
%\setcounter{equation}{0}
%%%%%%%%%%%%%%%%%%%%%%%%%%%%%%%%%%%%%%%%%%%%%%%%%%
In the weak-field limit, the gravitational field can be described as a linearized perturbation on a flat background spacetime, so the spacetime metric $g_{\mu\nu}$ can  accordingly be expanded as $g_{\mu\nu}=\eta_{\mu\nu}+h_{\mu\nu}$, where $\eta_{\mu\nu}$ is the flat spacetime metric, and $h_{\mu\nu}$ is the linear perturbation. 
Under this approximation, the interaction Lagrangian density between the gravitational fields and a nonpointlike object takes the standard form~\cite{Dyson1969,Weinberg}
\beq \label{Lagrangian density}
\mathcal{L}=\frac{1}{2}h_{\mu\nu}T^{\mu\nu},
\eeq
where $T^{\mu\nu}$ is the energy-momentum tensor of the object.  
We treat $h_{\mu\nu}$ and $ p_{\mu \nu}=\frac{\partial \mathcal{L}}{\partial \dot{h}_{\mu \nu}}$ as the generalized coordinate and the generalized momentum respectively, where $\dot{h}_{\mu \nu}$ is the generalized velocity, and the dot denotes derivative with respect to time $t$. 
To obtain the Hamiltonian density, the standard choice is to work in a local inertial frame, in which the generalized velocities $\dot{h}_{\mu \nu}$ are considered negligible. Therefore, the interaction Hamiltonian density corresponding to the above Lagrangian density can be expressed as~\cite{Boughn2006,Oniga2016}
\beq \label{gen_Ham}
\mathcal{H} =\frac{\partial \mathcal{L}}{\partial \dot{h}_{\mu \nu}}\dot{h}_{\mu \nu}-\mathcal{L}=-\frac{1}{2}h_{\mu\nu}T^{\mu\nu}.
\eeq
In the local inertial frame, the dominant term of the energy-momentum tensor is  $T^{00}=\rho_{m}$, where $\rho_{m}$ is the mass-energy density. In addition, we also include the $T^{0i}$ terms, where $T^{0i}=\rho_{m}v^{i}$ is the localized mass-current density. Then, the interaction Hamiltonian density can be expressed as
\bea
    \label{Hamiltonian density}
\nonumber
    \mathcal{H}&=&-\frac{1}{2}h_{00}T^{00}-\frac{1}{2}h_{0i}T^{0i}  \\
&=&-\frac{1}{2}h_{00}\rho_{m}(x)-\frac{1}{2}h_{0i} \rho_{m}(x)v^{i}.
\eea

Since the duration of the interaction between the fluctuating gravitational fields and the objects may be long, it is necessary to adopt a coordinate system that is locally inertial for an extended time~\cite{Boughn2006}. A suitable choice is to work in the Fermi normal coordinate system.
Here, we aim to establish a relation between the linear perturbation of the metric $h_{\mu\nu}$ and the corresponding Riemann tensor $R_{\mu\nu\alpha\beta}$. To this end, we need to express $h_{\mu\nu}$ as a Taylor expansion in powers of the Fermi coordinates. Note that the time component of Fermi coordinates is a constant~\cite{MTW,Manasse1963}, so the expansion terms dependent on the time coordinate naturally disappear, and the metric can be expanded as follows:
\begin{equation}
    \label{Fermi_2}
    g_{\mu \nu}=\eta _{\mu \nu}+\frac{1}{2}g_{\mu \nu ,ij}x^ix^j
    +O\left(x^3 \right),
\end{equation}
where $x^{i}$ denotes a spatial coordinate in the Fermi normal coordinate system. Replacing the second derivative of the metric tensor $g_{\mu \nu ,ij}$ in the equation above with the Riemann curvature tensor $R_{\alpha\beta\gamma\sigma}$, one obtains~\cite{Mashhoon2001,Mashhoon_review,Bini2022,Mashhoon2021(2),MTW}
\beq \label{h00}
h_{00}=-R_{0j0k}x^{j}x^{k},
\eeq and
\beq \label{h0i}
h_{0i}=-\frac{2}{3}R_{0jik}x^{j}x^{k}.
\eeq
Therefore, the dependence on the time enters the perturbation metric only through the components of the Riemann curvature tensor.  
Taking  Eqs. (\ref{h00}) and (\ref{h0i}) into Eq. \eqref{Hamiltonian density}, the interaction Hamiltonian density can further be written as
\bea
\label{new_Hamiltonian}
%\nonumber 
\mathcal{H}
&=&\frac{1}{2}\rho_{m}(x)x^{j}x^{k}R_{0j0k}+\frac{1}{3}R_{0jik}x^{j}x^{k}\rho_{m}(x)v^{i}.
\eea
According to the Weyl gravitoelectromagnetism, a gravitoelectric field,
\beq \label{my_Eij}
E_{ij}=-C_{0i0j},
\eeq
and a gravitomagnetic field,
\beq \label{Bij}
B_{ij}=\frac{1}{2}\epsilon_{ifl}C_{fl0j},
\eeq
can be defined by an analogy between the Maxwell equations and the linearized gravitational field equations~\cite{Campbell1976,Matte1953,Campbell1971,Szekeres1971,Maartens1998,Ruggiero2002,Ramos2010}, where $\epsilon_{ifl}$ is the spatial Levi-Civita tensor, and $C_{\alpha \beta \mu \nu}$ is the traceless part of the Riemann curvature tensor, i.e., the Weyl  tensor. In the vacuum case, the energy-momentum tensor $T_{\mu\nu}$ is zero, and the definitions in Eqs. \eqref{my_Eij} and \eqref{Bij} are equivalent to $E_{ij}=-R_{0i0j}$ and $B_{ij}=\frac{1}{2}\epsilon_{ifl}R_{fl0j}$. 
Then, the Hamiltonian density $\mathcal{H}$ can be rewritten as
\bea
    \label{Hamiltonian rewritten}
%\nonumber
\mathcal{H}
&=&-\frac{1}{2}\rho_{m}(x)x^{j}x^{k}E_{jk}-\frac{1}{3}\rho_{m}(x)\left({\textbf x}\times{\textbf v}\right)^{l}x^{j}B_{lj}. 
\eea
Allowing for the fact that $E_{ij}$ and $B_{ij}$ are symmetric traceless tensors, the interaction Hamiltonian can be expressed as
\bea
    \label{tot_Hamiltonian}
%\nonumber
    H&=&\int d^{3}x\mathcal{H}=
-\frac{1}{2}Q^{jk}E_{jk}-\frac{1}{3}S^{lj}B_{lj},
\eea
where $Q^{jk}\equiv\int d^{3}x\rho_{m}(x)(x^{j}x^{k}-\frac{1}{3}\delta^{jk}r^{2})$ is the mass quadrupole moment tensor, and $S^{lj}\equiv\frac{1}{2}\int d^{3}x\rho_{m}(x)[(\textbf{x}\times\textbf{v})^{l}x^{j}+(\textbf{x}\times\textbf{v})^{j}x^{l}]$ is the mass-current quadrupole moment of the object which has localized mass-current density~\cite{Flanagan2007,Rezzolla1999}.  So, the interaction Hamiltonian (\ref{tot_Hamiltonian}) represents the interaction between the mass quadrupole moment and the mass-current quadrupole moment of the object and the fluctuating gravitational fields. Note that the Hamiltonian (\ref{tot_Hamiltonian}) is gauge invariant. In contrast,  $\mathcal{L}$ is not gauge invariant, although it is a scalar in its mathematical form.

\section{The quantum gravitomagnetic interaction}\label{sec3}

The system we consider consists of a pair of gravitationally polarizable objects (labeled as A and B respectively), which are coupled with fluctuating gravitational fields in vacuum. The two objects are considered as two-level systems. The excited and ground states of the two objects are labeled as $|e_{A(B)}\rangle$ and $|g_{A(B)}\rangle$ respectively, and the energy level spacing is $\omega_{A(B)}$. Hence, the total Hamiltonian of this system can be expressed as
\beq \label{tot_sys_Hamiltonian}
H_{\rm tot}=H_{A}+H_{B}+H_{F}+H_{\rm int},
\eeq
where $H_{A(B)}$ is the Hamiltonian of object A(B), $H_{F}$ is the Hamiltonian of the gravitational fields, and $H_{\rm int}$ is the interaction Hamiltonian between the objects and the gravitational fields, which can be expressed in the form of Eq. (\ref{tot_Hamiltonian}) as
\beq \label{H_int}
H_{\rm int}=H^{\rm GE}_{\rm int}+H^{\rm GM}_{\rm int},
\eeq
where
\beq \label{H_GE}
H^{\rm GE}_{\rm int}\equiv-\frac{1}{2}Q^{ij}_{A}E_{ij}(\textbf{r}_{A})-\frac{1}{2}Q^{ij}_{B}E_{ij}(\textbf{r}_{B}),
\eeq
and
\beq \label{H_GM}
H^{\rm GM}_{\rm int}\equiv-\frac{1}{3}S^{ij}_{A}B_{ij}(\textbf{r}_{A})-\frac{1}{3}S^{ij}_{B}B_{ij}(\textbf{r}_{B}).
\eeq
As has been mentioned, the gravitational Casimir-Polder interaction studied in the existing literature~\cite{Wu2016,Wu2017,Ford2016,Holstein2017,yu2018,Hu2017,yongs2020epjc,yongs2022prd} originates from the interaction between the instantaneous mass quadrupole moments induced by the fluctuating gravitoelectric fields, which corresponds to the interaction Hamiltonian equation (\ref{H_GE}). In what follows, we study the contribution of the fluctuating gravitomagnetic fields to the quantum gravitational interaction between two nonpointlike objects. Correspondingly, the interaction Hamiltonian is given by Eq. (\ref{H_GM}).

In the transverse tracefree (TT) gauge, the linear gravitational perturbation can be quantized as~\cite{Wu2017,Oniga2016}
\beq \label{hij}
h_{ij}(\textbf{r},t)
=\sum_{\textbf{k},\lambda} \sqrt{\frac{8\pi G}{(2\pi)^3\omega}} \left[a_{\lambda}(\omega) e_{ij}(\textbf{k},\lambda)e^{i\textbf{k}\cdot\textbf{r}-i\omega t}+a^{\dagger}_{\lambda}(\omega) e_{ij}(\textbf{k},\lambda)e^{-i\textbf{k}\cdot\textbf{r}+i\omega t}\right],
\eeq
where $G$ is the Newton's gravitational constant, $\textbf{k}$  the wave vector, $\omega=|\textbf{k}|$  the frequency, $a_{\lambda}$ and $a^{\dagger}_{\lambda}$ the annihilation and creation operators, $e_{ij}(\textbf{k},\lambda)$ the polarization tensor, and $\lambda$ labels the polarization. In the weak-field approximation, the Riemann curvature tensor $R_{\alpha\beta\mu\nu}$ can be expressed in terms of the gravitational metric perturbation  as
\beq \label{R_tensor}
R_{\alpha\beta\mu\nu}=\frac{1}{2}\left(\partial_{\beta}\partial_{\mu}h_{\alpha\nu}-\partial_{\alpha}\partial_{\mu}h_{\beta\nu}-\partial_{\beta}\partial_{\nu}h_{\alpha\mu}+\partial_{\alpha}\partial_{\nu}h_{\beta\mu}\right).
\eeq
Then, in the TT gauge, the gravitomagnetic tensor can be written according to Eq. (\ref{Bij}) as 
\bea \label{Bij_h}
%\nonumber
B_{ij}
=-\frac{1}{2}\epsilon_{ifl}\partial_{f}\dot{h}_{lj},
\eea
where a dot represents the first derivative with respect to time $t$. Taking  Eq.~(\ref{hij}) into Eq.~(\ref{Bij_h}), the quantized gravitomagnetic field  can be obtained  as
\beq \label{quan_B_ij}
B_{ij}(\textbf{r},t)=\frac{i}{2}\sum_{\lambda}\int d^{3}\textbf{k}\sqrt{\frac{8\pi G\omega}{(2\pi)^{3}}}\epsilon_{ifl}\partial_{f}\left[a_{\lambda}(\omega,t) e_{lj}(\textbf{k},\lambda)e^{i\textbf{k}\cdot\textbf{r}}-a^{\dagger}_{\lambda}(\omega,t) e_{lj}(\textbf{k},\lambda)e^{-i\textbf{k}\cdot\textbf{r}}\right].
\eeq
Now we introduce a vector $\textbf{e}_{3}\equiv\frac{\textbf{k}}{|\textbf{k}|}$, which is the unit vector along the propagation direction of the gravitational field. Then, we have
\bea \label{Bij_h_further}
%\nonumber
B_{ij}(\textbf{r},t)
&=&-\frac{1}{2}\sum_{\lambda}\int d^{3}\textbf{k}\sqrt{\frac{8\pi G\omega^{3}}{(2\pi)^{3}}}\epsilon_{ifl}e^{f}_{3}e_{lj}(\textbf{k},\lambda)\left[a_{\lambda}(\omega,t)e^{i\textbf{k}\cdot\textbf{r}}+a^{\dagger}_{\lambda}(\omega,t)e^{-i\textbf{k}\cdot\textbf{r}}\right], 
\eea
where $e^{f}_{3}~(f=x,y,z)$ represents the $f$th coordinate component of the vector $\textbf{e}_{3}$.

We assume that the two objects are in their ground states. Since the coupling between the object and the gravitational field is linear in the object and field operators, each object should interact with the gravitational field at least twice and then return to its initial ground state. Therefore, the quantum gravitomagnetic interaction energy between two ground-state objects can be calculated by the fourth-order perturbation theory, which takes the standard form
\beq \label{E_AB}
\Delta E^{\rm GM}_{AB}=-{\sum_{\mathrm{I},\mathrm{II},\mathrm{III}}}'\frac{\langle \phi|H^{\rm GM}_{\rm int}|\mathrm{I}\rangle\langle \mathrm{I}|H^{\rm GM}_{\rm int}|\mathrm{II}\rangle\langle \mathrm{II}|H^{\rm GM}_{\rm int}|\mathrm{III}\rangle\langle \mathrm{III}|H^{\rm GM}_{\rm int}|\phi\rangle}{(E_{\mathrm{I}}-E_{\phi})(E_{\mathrm{II}}-E_{\phi})(E_{\mathrm{III}}-E_{\phi})}.
\eeq
Here $|\phi\rangle=|g_{A}\rangle|g_{B}\rangle|0\rangle$ is the ground state of the whole system, where $|0\rangle$ is the vacuum state of the fluctuating gravitational field. The primed summation means that $|\phi\rangle $ is excluded in the summation. Here, $| \mathrm{I}\rangle$, $| \mathrm{II}\rangle$ and $| \mathrm{III} \rangle$ are the three intermediate states in the interaction processes. 
During each interaction between the objects and the gravitational field, a virtual graviton may be emitted or absorbed by an object. Hence, the intermediate states $|\mathrm{I}\rangle$ and $|\mathrm{III}\rangle$ which are adjacent to the initial and final states respectively in Eq. \eqref{E_AB} must consist of a virtual graviton and an object in an excited state. For the intermediate state $|\mathrm{II}\rangle$, there are three possibilities, which can be summarized as (a) Both of the two objects are in the ground state and there are two virtual gravitons; (b) Both of the two objects are in the excited state and there are no virtual gravitons; (c) Both of the two objects are in the excited state and there are two virtual gravitons. See Table \ref{24I} in Appendix \ref{Appendix1} for the possible intermediate states and the corresponding denominators.

Summing up all the contributions of possible intermediate states, we obtain the quantum gravitomagnetic interaction energy between the pair of objects as
\bea \label{E_AB_int}
\nonumber
\Delta E^{\rm GM}_{AB}(\textbf{r}_{A},\textbf{r}_{B})&=&-\frac{1}{81}\int_{0}^{+\infty}d\omega \int_{0}^{+\infty}d{\omega}'\sum\limits_{n=1}^{12}\frac{1}{D_{n}}\hat{S}^{ij}_{A}\hat{S}^{*kl}_{A}\hat{S}^{ab}_{B}\hat{S}^{*cd}_{B}  
\\
\nonumber&&\times G_{ijab}(\omega,\textbf{r}_{A},\textbf{r}_{B})G_{klcd}({\omega}',\textbf{r}_{A},\textbf{r}_{B}) \\
\nonumber&=&-\frac{1}{81}\int_{0}^{+\infty}d\omega \int_{0}^{+\infty}d{\omega}'\hat{S}^{ij}_{A}\hat{S}^{*kl}_{A}\hat{S}^{ab}_{B}\hat{S}^{*cd}_{B} G_{ijab}(\omega,\textbf{r}_{A},\textbf{r}_{B})G_{klcd}({\omega}',\textbf{r}_{A},\textbf{r}_{B})  \\
&&\times\frac{4\left(\omega_{A}+\omega_{B}+\omega\right)}{\left(\omega_{A}+\omega_{B}\right)\left(\omega_{A}+\omega)(\omega_{B}+\omega\right)}\left(\frac{1}{\omega+{\omega}'}-\frac{1}{\omega-{\omega}'}\right).
\eea
Here $D_{n}$ $(n=1,2,3,...,12)$ are the energy denominators in Eq.~(\ref{E_AB}) shown in Table~\ref{24I}. $\hat{S}^{ij}_{A(B)}=\langle g_{A(B)}|S^{ij}_{A(B)}|e_{A(B)}\rangle$ is the quadrupole transition matrix element, and $\hat{S}^{*ij}_{A(B)}=\langle e_{A(B)}|S^{ij}_{A(B)}|g_{A(B)}\rangle$ is the corresponding conjugate term. 
$G_{ijab}(\omega, \textbf{r}_{A},\textbf{r}_{B})$ is the two-point correlation function of the gravitomagnetic fields in the frequency domain, which takes the form
\beq \label{G_ijab}
G_{ijab}(\omega, \textbf{r}_{A},\textbf{r}_{B})=\langle 0|B_{ij}(\omega,\textbf{r}_{A})B_{ab}(\omega,\textbf{r}_{B})|0\rangle.
\eeq
We assume that the objects are isotropically polarizable, then the relation satisfied by the product between the quadrupole transition matrix element and its conjugate can be expressed as
\beq \label{polar_1}
\hat{S}^{ij}_{A(B)}\hat{S}^{*kl}_{A(B)}=(\delta_{ik}\delta_{jl}+\delta_{il}\delta_{jk})\hat{\chi}_{A(B)},\\
\eeq
where  $\hat{\chi}\equiv|\hat{S}^{ij}|^{2}$. Substituting Eq. (\ref{polar_1}) into Eq. (\ref{E_AB_int}),  one  obtains
\bea \label{re_E_AB_int}
\nonumber
\Delta E^{\rm GM}_{AB}(\textbf{r}_{A},\textbf{r}_{B})&=&-\frac{4}{81(\omega_{A}+\omega_{B})}\int_{0}^{+\infty}d\omega \int_{0}^{+\infty}d{\omega}' G_{ijab}(\omega,\textbf{r}_{A},\textbf{r}_{B})G_{ijab}({\omega}',\textbf{r}_{A},\textbf{r}_{B})  \\
&&\times\frac{\hat{\chi}_{A}\hat{\chi}_{B}\left(\omega_{A}+\omega_{B}+\omega\right)}{\left(\omega_{A}+\omega)(\omega_{B}+\omega\right)}\left(\frac{1}{\omega+{\omega}'}-\frac{1}{\omega-{\omega}'}\right).
\eea
The two-point  function $G_{ijab}(\omega,\textbf{r}_{A},\textbf{r}_{B})$ can be obtained from $G_{ijab}(\textbf{r}_{A},\textbf{r}_{B},t_{A},t_{B})$ by the Fourier transforms. In the time domain, the two-point  function of the gravitomagnetic field can be obtained by using  Eq. (\ref{Bij_h_further}) as
\bea \label{G_ijab_cal}
\nonumber
G_{ijab}(\textbf{r},{\textbf{r}}',t,{t}')&=&\langle 0|B_{ij}(\textbf{r},t)B_{ab}(\textbf{r}',{t}')|0\rangle  \\
&=&\int d^{3}\textbf{k}\frac{G\omega^{3}}{(2\pi)^{2}}\mathcal{G}_{ijab}(\textbf{k})e^{i\textbf{k}\cdot(\textbf{r}-{\textbf{r}}')-i\omega(t-{t}')},  
\eea
where $\mathcal{G}_{ijab}(\textbf{k})$ in the  equation above is the polarization summation term,  which can be expressed as
\bea \label{polar_sum}
\nonumber
\mathcal{G}_{ijab}(\textbf{k})&=&\epsilon_{ifl}e^{f}_{3}\epsilon_{apq}e^{p}_{3}\sum_{\lambda}e_{lj}(\textbf{k},\lambda)e_{qb}(\textbf{k},\lambda)=\sum_{\lambda}e_{ij}(\textbf{k},\lambda)e_{ab}(\textbf{k},\lambda) \\
\nonumber&=&\delta_{ia}\delta_{jb}+\delta_{ib}\delta_{ja}-\delta_{ij}\delta_{ab}+\hat{k}_{i}\hat{k}_{j}\hat{k}_{a}\hat{k}_{b}+\hat{k}_{i}\hat{k}_{j}\delta_{ab}+\hat{k}_{a}\hat{k}_{b}\delta_{ij}-\hat{k}_{i}\hat{k}_{a}\delta_{jb}-\hat{k}_{i}\hat{k}_{b}\delta_{ja}  \\  
&&-\hat{k}_{j}\hat{k}_{a}\delta_{ib}-\hat{k}_{j}\hat{k}_{b}\delta_{ia},
\eea
with $\hat{k}_{i}$ being the $i$th component of the unit vector  $\hat{k}=\textbf{k}/k$. The derivation of Eq. (\ref{polar_sum}) is shown in  Appendix~\ref{appd3}.
Transforming to the spherical coordinate, i.e., letting $\hat{k}_{x}=\sin\theta\cos\varphi$, $\hat{k}_{y}=\sin\theta\sin\varphi$ and $\hat{k}_{z}=\cos\theta$, and labeling
\beq \label{g_ijab}
\mathcal{G}_{ijab}(\textbf{k})\xrightarrow{(\theta,\varphi)}\mathcal{G}_{ijab}(\theta,\varphi),
\eeq
 the two-point correlation function in the time domain can be written as 
\beq \label{G_ijab_sph}
G_{ijab}(r,\Delta t)=\int_{0}^{+\infty}d\omega\frac{G\omega^{5}}{(2\pi)^{2}}\int_{0}^{\pi}d\theta\sin\theta\int_{0}^{2\pi}d\varphi \mathcal{G}_{ijab}(\theta,\varphi)e^{i\omega(r\cos\theta-\Delta t)},
\eeq
where $r=|\textbf{r}-{\textbf{r}}'|$, and $\Delta t=t-{t}'$. Performing the Fourier transform, one obtains the two-point correlation function in the frequency domain as
\bea \label{G_ijab_fre}
\nonumber
G_{ijab}(\tilde{\omega},\textbf{r}_{A},\textbf{r}_{B})&=&\frac{1}{2\pi}\int_{-\infty}^{+\infty}d(\Delta t)e^{i\tilde{\omega}\Delta t}G_{ijab}(r,\Delta t)  \\
&=&\frac{G\tilde{\omega}^{5}}{(2\pi)^{2}}\int_{0}^{\pi}d\theta\sin\theta\int_{0}^{2\pi}d\varphi \mathcal{G}_{ijab}(\theta,\varphi)e^{i\tilde{\omega} r\cos\theta}.
\eea
Plugging Eq. (\ref{G_ijab_fre}) into  Eq. (\ref{re_E_AB_int}), and performing the integral over  $(\theta,\varphi,{\theta}’,{\varphi}’)$, we obtain
\bea \label{E_AB_r}
\nonumber
\Delta E^{\rm GM}_{AB}(r)&=&-\frac{32G^{2}}{81\pi^{2}(\omega_{A}+\omega_{B})r^{10}}\int_{0}^{+\infty}d\omega\int_{0}^{+\infty}d{\omega}'\frac{\hat{\chi}_{A}\hat{\chi}_{B}(\omega_{A}+\omega_{B}+\omega)}{(\omega_{A}+\omega)(\omega_{B}+\omega)}       \\
&&\times\left(\frac{1}{\omega+{\omega}'}-\frac{1}{\omega-{\omega}'}\right)\Big[F_{1}(\omega r,{\omega}'r)\cos({\omega}'r)+F_{2}(\omega r,{\omega}'r)\sin({\omega}'r)\Big],
\eea
where 
\bea \label{F_1}
\nonumber
F_{1}(\omega r,{\omega}'r)&=&\left(\omega r\right)\left({\omega}'r\right)\left[315+8(\omega r)^{2}({\omega}' r)^{2}-30(\omega r)^{2}-30({\omega}' r)^{2}\right]\cos(\omega r)       \\
\nonumber&&-({\omega}'r)\Big[315-135(\omega r)^{2}-30({\omega}'r)^{2}+18(\omega r)^{2}({\omega}'r)^{2}+3(\omega r)^{4}-2(\omega r)^{4}  \\
&&\times({\omega}'r)^{2}\Big]\sin(\omega r), 
\eea
and
\bea \label{F_2}
\nonumber
F_{2}(\omega r,{\omega}'r)&=&(\omega r)\left[-315+135({\omega}' r)^{2}+30(\omega r)^{2}-18(\omega r)^{2}({\omega}'r)^{2}-3({\omega}' r)^{4}+2(\omega r)^{2}({\omega}'r)^{4}\right]  \\
\nonumber&&\times\cos(\omega r)+\Big[315-135(\omega r)^{2}+3(\omega r)^{4}-135({\omega}'r)^{2}+63(\omega r)^{2}({\omega}'r)^{2}     \\
&&-3(\omega r)^{4}({\omega}'r)^{2}+3({\omega }'r)^{4}-3(\omega r)^{2}({\omega}'r)^{4}+(\omega r)^{4}({\omega}'r)^{4}\Big]\sin(\omega r).
\eea
Obviously, $F_{1}(\omega r,-{\omega}'r)=-F_{1}(\omega r,{\omega}'r)$ 
and $F_{2}(\omega r,-{\omega}'r)=F_{2}(\omega r,{\omega}'r)$. 
So, the gravitomagnetic interaction energy $\Delta E^{\rm GM}_{AB}(r)$ can further be written as
\bea \label{E_AB_ex}
%\nonumber
\Delta E^{\rm GM}_{AB}(r)
\nonumber&=&-\frac{16G^{2}}{81\pi^{2}(\omega_{A}+\omega_{B})r^{10}}\int_{0}^{+\infty}d\omega\frac{\hat{\chi}_{A}\hat{\chi}_{B}(\omega_{A}+\omega_{B}+\omega)}{(\omega_{A}+\omega)(\omega_{B}+\omega)}  \\
&&\times\int_{-\infty}^{+\infty}d{\omega}'\left(\frac{1}{\omega+{\omega}'}+\frac{1}{-\omega+{\omega}'}\right)\Big[F_{1}(\omega r,{\omega}'r)-iF_{2}(\omega r,{\omega}'r)\Big]e^{i{\omega}'r}.
\eea
Performing the principle value integral on ${\omega}'$ in the equation above gives
\bea \label{E_AB_prin_int}
\nonumber
\Delta E^{\rm GM}_{AB}(r)&=&-\frac{16G^{2}}{81\pi(\omega_{A}+\omega_{B})r^{10}}\int_{0}^{+\infty}d\omega\frac{\hat{\chi}_{A}\hat{\chi}_{B}(\omega_{A}+\omega_{B}+\omega)}{(\omega_{A}+\omega)(\omega_{B}+\omega)}   \\
&&\times\Big[F_{3}(\omega r)\cos(2\omega r)+F_{4}(\omega r)\sin(2\omega r)\Big],
\eea
where
\beq \label{F_3}
F_{3}(\omega r)=-630(\omega r)+330(\omega r)^{3}-42(\omega r)^{5}+4(\omega r)^{7},
\eeq
\beq \label{F_4}
F_{4}(\omega r)=315-585(\omega r)^{2}+129(\omega r)^{4}-14(\omega r)^{6}+(\omega r)^{8}.
\eeq
Note that $F_{3}(-\omega r)=-F_{3}(\omega r)$ and $F_{4}(-\omega r)=F_{4}(\omega r)$, and as  a result,  Eq. (\ref{E_AB_prin_int}) can further be  expressed as
\bea \label{E_AB_F_34}
%\nonumber
\Delta E^{\rm GM}_{AB}(r)
\nonumber&=&-\frac{8G^{2}}{81\pi(\omega_{A}+\omega_{B})r^{10}}
\Bigg\{\int_{0}^{+\infty}d\omega\frac{\hat{\chi}_{A}\hat{\chi}_{B}(\omega_{A}+\omega_{B}+\omega)}{(\omega_{A}+\omega)(\omega_{B}+\omega)}\Big[F_{3}(\omega r)-iF_{4}(\omega r)\Big] e^{i2\omega r}   \\
&&+\int_{0}^{-\infty}d\omega\frac{\hat{\chi}_{A}\hat{\chi}_{B}(\omega_{A}+\omega_{B}-\omega)}{(\omega_{A}-\omega)(\omega_{B}-\omega)}\Big[F_{3}(\omega r)-iF_{4}(\omega r)\Big]e^{i2\omega r}\Bigg\}.
\eea
Letting $\omega=iu$, and performing the  integral above on the imaginary axis, one obtains 
\beq \label{E_AB_F_34_simpl_end}
\Delta E^{\rm GM}_{AB}(r)=-\frac{16G^{2}}{81\pi r^{10}}\int_{0}^{+\infty}du\chi_{A}(iu)\chi_{B}(iu)T(ur)e^{-2ur},
\eeq
where
\bea \label{T_ur}
T(x)=315+630x+585x^{2}+330x^{3}+129x^{4}+42x^{5}+14x^{6}+4x^{7}+x^{8},
\eea
and
\beq \label{chi_A}
\chi_{A(B)}(iu)=\lim\limits_{\epsilon\to0^{+}}\frac{\hat{\chi}_{A(B)}\omega_{A(B)}}{\omega^{2}_{A(B)}-(iu)^{2}-i\epsilon(iu)}
\eeq
is defined as the object's ground-state gravitomagnetic polarizability satisfying
\beq \label{chi_A(B)}
S_{ij}(iu)=\chi(iu)B_{ij}(iu,\textbf{r}).
\eeq

Now, let us discuss the asymptotic behaviors of the quantum gravitomagnetic interaction energy \eqref{E_AB_F_34_simpl_end} in the near and far regimes respectively. In the near regime, i.e., when the distance between the two objects $r$ is much smaller than the transition wavelength of the objects $\omega^{-1}_{A(B)}$, all the terms in the integrand containing $ur$ can be neglected, so the quantum gravitomagnetic interaction energy takes the form
\beq \label{E_AB, near_SI}
\Delta E^{\rm GM, near}_{AB}(r)=-\frac{560\hbar G^{2}}{9\pi r^{10}}\int_{0}^{+\infty}du\chi_{A}(iu)\chi_{B}(iu).
\eeq
Note that here the result is shown in the SI units (International System of Units). According to the definition of the gravitomagnetic polarizability in Eq.~(\ref{chi_A}), the frequency dependence of the gravitomagnetic polarizability $\chi_{A(B)}(\omega)$ can be expressed as
\beq \label{fre-dep-polar}
 \chi _{A(B)}(\omega) =\frac{\chi _{A(B)}(0)}{1-\left( \frac{\omega}{\omega _{A(B)}} \right) ^2}.
\eeq
Substituting Eq.~(\ref{fre-dep-polar}) into Eq.~(\ref{E_AB, near_SI}), and performing the integration with respect to the variable $u$, we can further obtain an explicit expression for the interaction energy in the near regime as
\beq \label{near interaction}
 \varDelta E_{AB}^{\rm GM,near}\left( r \right)=-\frac{280\hbar G^2}{9 r^{10}}\frac{\omega _A\omega _B}{\omega _A+\omega _B}\chi _A\left( 0 \right) \chi _B\left( 0 \right).
\eeq
Note that $\chi_ {A (B)}(0)$ represents the static gravitomagnetic polarizability of object $A(B)$. 
On the other hand, in the far regime, i.e., when the distance between the two objects $r$ is much larger than the transition wavelength of the objects $\omega^{-1}_{A(B)}$, the frequency-dependent polarizability $\chi_{A(B)}(iu)$ can be approximated with the static one $\chi_{A(B)}(0)$ due to the exponential decay term in Eq. (\ref{E_AB_F_34_simpl_end}). By using integration by parts, the quantum gravitomagnetic interaction energy in the far regime is found to be
\beq \label{E_AB, far_SI}
\Delta E^{\rm GM, far}_{AB}(r)=-\frac{1772\hbar c G^{2}}{9\pi r^{11}}\chi_{A}(0)\chi_{B}(0).
\eeq
By setting Eqs.~\eqref{near interaction} and \eqref{E_AB, far_SI} equal,  we can find the critical distance $r^*$ at which we transition from the near regime to the far regime:
\beq \label{critical distance}
    r^*=\frac{443c\left( \omega _A+\omega _B \right)}{70\pi \omega _A\omega _B},
\eeq 
which is of the order of $c/\omega_{A(B)}$ as expected. This critical distance marks the boundary where the nature of the interaction shifts, indicating the transition between different scaling behaviors in the gravitational interaction.

The calculations above show that the quantum gravitomagnetic interaction between two nonpointlike objects induced by fluctuating gravitomagnetic fields in vacuum shows an $r^{-10}$ dependence in the near regime and an $r^{-11}$ dependence in the far regime. A comparison between Eqs.~\eqref{E_AB, far_SI} and \eqref{E_AB, near_SI} and Eqs.~(1) and (2) in Ref. \cite{Ford2016} shows that, although the distance dependence is the same as that of the quantum gravitational interaction between two objects induced by fluctuating gravitoelectric fields, the coefficients are different. This contrasts significantly with the electromagnetic case, where the Casimir-Polder interaction due to the interaction between instantaneous magnetic dipoles is of the same form as that due to the interaction between instantaneous electric dipoles, apart from the replacement of electric polarizability with magnetic susceptibility~\cite{Buhmann2005,Buhmann_book,Salam2010}. 
Moreover,  the quantum gravitomagnetic interaction energy is proportional to $G^2$, which can be understood as a result of an exchange of two virtual gravitons according to the rules of the Feynman diagram. In this sense, it is of the same order as that of the quantum gravitational interaction induced by fluctuating gravitoelectric fields. 
However, according to the interaction Lagrangian density \eqref{Lagrangian density}, in the gravitomagnetic case, it is the mass-current density $\rho_{m}v^{i}$ rather than the mass-energy density  $\rho_{m}$ that is coupled to the fluctuating gravitational fields $h_{\mu\nu}$. Therefore, the ratio between the gravitomagnetic interaction Lagrangian and the gravitoelectric one is of the order of $v/c$, and correspondingly, the ratio between the quantum gravitomagnetic interaction energy and the quantum gravitoelectric interaction energy is of the order of $v^4/c^4$. This is why the contribution of the fluctuating gravitomagnetic fields to the quantum gravitational interaction energy was not considered in previous studies. Nevertheless, if there exists some specific material such that the induced mass-current velocity can be close to the speed of light, or equivalently, the gravitomagnetic polarizability is very large, then the contribution of the fluctuating gravitomagnetic fields to the quantum gravitational interaction energy should be non-negligible.

Finally, we would like to note that, when fluctuating gravitomagnetic fields are considered, there should be contribution to the quantum gravitational interaction from the gravitoelectric-gravitomagnetic cross terms, which is of the order of $v^2/c^2$. This  is in addition to the gravitoelectric-gravitoelectric interaction investigated in previous studies \cite{Wu2016,Ford2016,Holstein2017}, and the quantum gravitomagnetic-gravitomagnetic interaction investigated here. 
It is expected that, in the retarded regime, the distance dependence of the quantum gravitational interaction from the gravitoelectric-gravitomagnetic cross terms ($\propto r^{-11}$) is the same as that in the gravitoelectric-gravitoelectric and gravitomagnetic-gravitomagnetic cases. However, in the near regime, the distance dependence should be different. We hope to leave the explicit investigation of these interactions for future work.

\section{Summary }
\label{sec_disc}
%\setcounter{equation}{0}
%%%%%%%%%%%%%%%%%%%%%%%%%%%%%%%%%%%%%%%%%%%%%%%%%%

In summary, we have investigated the quantum gravitational interaction between two nonpointlike objects coupled with fluctuating gravitomagnetic fields in vacuum within the framework of linearized quantum gravity. We found that these fluctuating gravitomagnetic fields induce instantaneous localized mass currents in the objects, which interact to generate a quantum gravitomagnetic interaction energy. Using fourth-order perturbation theory, we determined that this interaction exhibits  an $r^{-10}$ dependence in the near regime and an $r^{-11}$ dependence in the far regime, where $r$ is the distance between the two objects. The contribution of the fluctuating gravitomagnetic fields to the quantum gravitational interaction between two objects is expected to be significant when the gravitomagnetic polarizability of the objects is large.

\begin{acknowledgments}

We would like to thank Puxun Wu and Yongshun Hu for helpful discussions. We would also like to thank the anonymous referees for their insightful comments and helpful suggestions. This work was supported in part by the NSFC under Grant No. 12075084, and the innovative research group of Hunan Province under Grant No. 2024JJ1006.

\end{acknowledgments}

%\newpage

\appendix
\section{Intermediate processes of the quantum gravitomagnetic interaction}\label{Appendix1}
The intermediate states and the associated energy denominators of Eq.~(\ref{E_AB}) are as follows.

\begin{table}[H]
  \centering
  \scalebox{0.75}{
\begin{tabular}{cllll}
\hline
Case  &\hspace{1ex} $|\text{I}\rangle$ &\hspace{2ex}$|\text{II}\rangle$ &\hspace{2ex}$|\text{III}\rangle$ &\hspace{2ex} Denominator\\
\hline
(1) &\hspace{1ex} $|e_A\rangle |g_B\rangle |1\rangle $
      & \hspace{2ex}$|g_A\rangle |g_B\rangle |1,{1}'\rangle $
      & \hspace{2ex}$|g_A\rangle |e_B\rangle |{1}'\rangle $
      & \hspace{2ex}$D_{1}=({\omega}'+\omega_{B})({\omega}'+\omega)(\omega+\omega_{A})$  \\
(2) &\hspace{1ex} $|e_A\rangle |g_B\rangle |1\rangle$
      & \hspace{2ex}$|g_A\rangle |g_B\rangle |1,{1}'\rangle$
      & \hspace{2ex}$|g_A\rangle |e_B\rangle |1\rangle$
      & \hspace{2ex}$D_{2}=(\omega+\omega_{B})({\omega}'+\omega)(\omega+\omega_{A})$  \\
(3) &\hspace{1ex} $|e_A\rangle |g_B\rangle |1\rangle$
      & \hspace{2ex}$|e_A\rangle |e_B\rangle |0\rangle $
      & \hspace{2ex}$|g_A\rangle |e_B\rangle |{1}'\rangle $
      & \hspace{2ex}$D_{3}=({\omega}'+\omega_{B})(\omega_{B}+\omega_{A})(\omega+\omega_{A})$  \\
(4) &\hspace{1ex} $|e_A\rangle |g_B\rangle |1\rangle$
      & \hspace{2ex}$|e_A\rangle |e_B\rangle |0\rangle$
      & \hspace{2ex}$|e_A\rangle |g_B\rangle |{1}'\rangle $
      & \hspace{2ex}$D_{4}=({\omega}'+\omega_{A})(\omega_{B}+\omega_{A})(\omega+\omega_{A})$  \\
(5) &\hspace{1ex} $|e_A\rangle |g_B\rangle |1\rangle$
      & \hspace{2ex}$|e_A\rangle |e_B\rangle |1,{1}'\rangle $
      & \hspace{2ex}$|g_A\rangle |e_B\rangle |1\rangle $
      & \hspace{2ex}$D_{5}=(\omega+\omega_{B})(\omega_{B}+\omega_{A}+{\omega}'+\omega)(\omega+\omega_{A})$  \\
(6) &\hspace{1ex} $|e_A\rangle |g_B\rangle |1\rangle$
      & \hspace{2ex}$|e_A\rangle |e_B\rangle |1,{1}'\rangle$
      & \hspace{2ex}$|e_A\rangle |g_B\rangle |{1}'\rangle $
      & \hspace{2ex}$D_{6}=({\omega}'+\omega_{A})(\omega_{B}+\omega_{A}+{\omega}'+\omega)(\omega+\omega_{A})$  \\
(7) &\hspace{1ex} $|g_A\rangle |e_B\rangle |1\rangle $
      & \hspace{2ex}$|g_A\rangle |g_B\rangle |1,{1}'\rangle $
      & \hspace{2ex}$|e_A\rangle |g_B\rangle |{1}'\rangle $
      & \hspace{2ex}$D_{7}=({\omega}'+\omega_{A})({\omega}'+\omega)(\omega+\omega_{B})$  \\
(8) &\hspace{1ex} $|g_A\rangle |e_B\rangle |1\rangle$
      & \hspace{2ex}$|g_A\rangle |g_B\rangle |1,{1}'\rangle$
      & \hspace{2ex}$|e_A\rangle |g_B\rangle |1\rangle$
      & \hspace{2ex}$D_{8}=(\omega+\omega_{A})({\omega}'+\omega)(\omega+\omega_{B})$  \\
(9) &\hspace{1ex} $|g_A\rangle |e_B\rangle |1\rangle$
      & \hspace{2ex}$|e_A\rangle |e_B\rangle |0\rangle $
      & \hspace{2ex}$|e_A\rangle |g_B\rangle |{1}'\rangle $
      & \hspace{2ex}$D_{9}=({\omega}'+\omega_{A})(\omega_{B}+\omega_{A})(\omega+\omega_{B})$  \\
(10)&\hspace{1ex} $|g_A\rangle |e_B\rangle |1\rangle$
      & \hspace{2ex}$|e_A\rangle |e_B\rangle |0\rangle$
      & \hspace{2ex}$|g_A\rangle |e_B\rangle |{1}'\rangle $
      & \hspace{2ex}$D_{10}=({\omega}'+\omega_{B})(\omega_{B}+\omega_{A})(\omega+\omega_{B})$  \\
(11)&\hspace{1ex} $|g_A\rangle |e_B\rangle |1\rangle$
      & \hspace{2ex}$|e_A\rangle |e_B\rangle |1,{1}'\rangle$
      & \hspace{2ex}$|e_A\rangle |g_B\rangle |1\rangle $
      & \hspace{2ex}$D_{11}=(\omega+\omega_{A})(\omega_{B}+\omega_{A}+{\omega}'+\omega)(\omega+\omega_{B})$  \\
(12)&\hspace{1ex} $|g_A\rangle |e_B\rangle |1\rangle$
      & \hspace{2ex}$|e_A\rangle |e_B\rangle |1,{1}'\rangle$
      & \hspace{2ex}$|g_A\rangle |e_B\rangle |{1}'\rangle$
      & \hspace{2ex}$D_{12}=({\omega}'+\omega_{B})(\omega_{B}+\omega_{A}+{\omega}'+\omega)(\omega+\omega_{B})$  \\
\hline
\end{tabular}}
  \caption{Twelve intermediate states contributing to the interaction energy and the explicit expressions of the corresponding energy denominators.}\label{24I}
\end{table}

\section{Summation of gravitomagnetic polarization tensors in the TT gauge}\label{appd3}
\begin{figure}
  \centering
   %Requires \usepackage{graphicx}
  \includegraphics[width=0.6\textwidth]{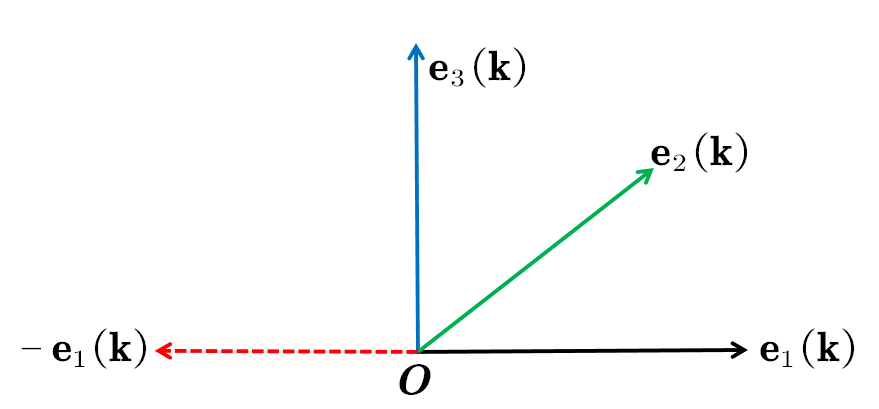}\\
  \caption{The schematic diagram of the triad of unit vectors and their orthogonal relations.}\label{T}
\end{figure}

As shown in Fig.~\ref{T}, we introduce a coordinate-independent triad of unit orthonormal vectors $[\textbf{e}_{1}(\textbf{k}),\textbf{e}_{2}(\textbf{k}),\textbf{e}_{3}(\textbf{k})]$. Here, $\textbf{e}_{3}(\textbf{k})=\textbf{k}/k\equiv\hat{k}$ is the unit vector in the direction of gravitational perturbation propagation.
The orthogonal relation satisfied by the triad can be written in the coordinate system describing the spacetime metric as~\cite{yu1999}
\beq \label{orth_ rela}
e^{i}_{a}(\textbf{k})e^{i}_{b}(\textbf{k})=\delta_{ab}, \quad a,b=1,2,3,
\eeq
and
\beq \label{orth_ 2}
e^{i}_{a}(\textbf{k})e^{j}_{a}(\textbf{k})=e^{i}_{1}e^{j}_{1}+e^{i}_{2}e^{j}_{2}+\hat{k}^{i}\hat{k}^{j}=\delta_{ij}, \quad i,j=x,y,z,
\eeq
where  $\hat{k}_{i}$ is the $i$th coordinate component of the unit vector $\hat{k}$. 
On the other hand, the cross product relations that this triad of unit vectors satisfies can be written in the coordinate system as
\beq \label{cro_ pro_1}
\textbf{e}_{3}(\textbf{k})\times\textbf{e}_{1}(\textbf{k})=\textbf{e}_{2}(\textbf{k})\Rightarrow\epsilon_{ijk}e^{j}_{3}e^{k}_{1}=e^{i}_{2},
\eeq
\beq \label{cro_ pro_2}
\textbf{e}_{3}(\textbf{k})\times\textbf{e}_{2}(\textbf{k})=-\textbf{e}_{1}(\textbf{k})\Rightarrow\epsilon_{ijk}e^{j}_{3}e^{k}_{2}=-e^{i}_{1}.
\eeq
Therefore, the gravitational polarization tensor $e_{ij}(\textbf{k},\lambda)$ in the TT gauge can be expressed by the vectors $\textbf{e}_{1}(\textbf{k})$ and $\textbf{e}_{2}(\textbf{k})$ in this triad as~\cite{MTW}
\beq \label{polar tensors_1}
e^{ij}(\textbf{k},+)=e^{i}_{1}(\textbf{k})\otimes e^{j}_{1}(\textbf{k})-e^{i}_{2}(\textbf{k})\otimes e^{j}_{2}(\textbf{k}),
\eeq
\beq \label{polar tensors_2}
e^{ij}(\textbf{k},\times)=e^{i}_{1}(\textbf{k})\otimes e^{j}_{2}(\textbf{k})+e^{i}_{2}(\textbf{k})\otimes e^{j}_{1}(\textbf{k}).
\eeq
Then we  obtain that
\bea \label{sum_polar tensors}
\nonumber
\sum_{\lambda}e_{ij}(\textbf{\textbf{k}},\lambda)e_{kl}(\textbf{k},\lambda)&=&e^{ij}(\textbf{k},+)e^{kl}({\textbf{k},+})+e^{ij}(\textbf{k},\times)e^{kl}(\textbf{k},\times)  \\
\nonumber&=&\left[e^{i}_{1}(\textbf{k})\otimes e^{j}_{1}(\textbf{k})-e^{i}_{2}(\textbf{k})\otimes e^{j}_{2}(\textbf{k})\right]\left[e^{k}_{1}(\textbf{k})\otimes e^{l}_{1}(\textbf{k})-e^{k}_{2}(\textbf{k})\otimes e^{l}_{2}(\textbf{k})\right]   \\
\nonumber&&+\left[e^{i}_{1}(\textbf{k})\otimes e^{j}_{2}(\textbf{k})+e^{i}_{2}(\textbf{k})\otimes e^{j}_{1}(\textbf{k})\right]\left[e^{k}_{1}(\textbf{k})\otimes e^{l}_{2}(\textbf{k})+e^{k}_{2}(\textbf{k})\otimes e^{l}_{1}(\textbf{k})\right].  \\
\eea
Using  Eq. (\ref{sum_polar tensors}), the gravitomagnetic polarization summation term  labeled as $\mathcal{G}_{ijab}(\textbf{k})$ in Eq.~(\ref{G_ijab_cal}) can be further expressed as
\bea \label{re_polar_sum}
\nonumber
\mathcal{G}_{ijab}(\textbf{k}) &=&\epsilon_{ifl}e^{f}_{3}\epsilon_{apq}e^{p}_{3}\Big\{\left[e^{l}_{1}(\textbf{k})\otimes e^{j}_{1}(\textbf{k})-e^{l}_{2}(\textbf{k})\otimes e^{j}_{2}(\textbf{k})\right]\left[e^{q}_{1}(\textbf{k})\otimes e^{b}_{1}(\textbf{k})-e^{q}_{2}(\textbf{k})\otimes e^{b}_{2}(\textbf{k})\right]  \\
\nonumber&&+\left[e^{l}_{1}(\textbf{k})\otimes e^{j}_{2}(\textbf{k})+e^{l}_{2}(\textbf{k})\otimes e^{j}_{1}(\textbf{k})\right]\left[e^{q}_{1}(\textbf{k})\otimes e^{b}_{2}(\textbf{k})+e^{q}_{2}(\textbf{k})\otimes e^{b}_{1}(\textbf{k})\right]\Big\}  \\
\nonumber&=&\epsilon_{ifl}e^{f}_{3}\left[e^{l}_{1}(\textbf{k})\otimes e^{j}_{1}(\textbf{k})-e^{l}_{2}(\textbf{k})\otimes e^{j}_{2}(\textbf{k})\right]\epsilon_{apq}e^{p}_{3}\left[e^{q}_{1}(\textbf{k})\otimes e^{b}_{1}(\textbf{k})-e^{q}_{2}(\textbf{k})\otimes e^{b}_{2}(\textbf{k})\right]  \\
\nonumber&&+\epsilon_{ifl}e^{f}_{3}\left[e^{l}_{1}(\textbf{k})\otimes e^{j}_{2}(\textbf{k})+e^{l}_{2}(\textbf{k})\otimes e^{j}_{1}(\textbf{k})\right]\epsilon_{apq}e^{p}_{3}\left[e^{q}_{1}(\textbf{k})\otimes e^{b}_{2}(\textbf{k})+e^{q}_{2}(\textbf{k})\otimes e^{b}_{1}(\textbf{k})\right]    \\
\nonumber&=&\Big\{\left[\epsilon_{ifl}e^{f}_{3}(\textbf{k})e^{l}_{1}(\textbf{k})\right]\otimes e^{j}_{1}(\textbf{k})-\left[\epsilon_{ifl}e^{f}_{3}(\textbf{k})e^{l}_{2}(\textbf{k})\right]\otimes e^{j}_{2}(\textbf{k})\Big\} \\
\nonumber&&\times\Big\{\left[\epsilon_{apq}e^{p}_{3}(\textbf{k})e^{q}_{1}(\textbf{k})\right]\otimes e^{b}_{1}(\textbf{k})-\left[\epsilon_{apq}e^{p}_{3}(\textbf{k})e^{q}_{2}(\textbf{k})\right]\otimes e^{b}_{2}(\textbf{k})\Big\}  \\
\nonumber&&+\Big\{\left[\epsilon_{ifl}e^{f}_{3}(\textbf{k})e^{l}_{1}(\textbf{k})\right]\otimes e^{j}_{2}(\textbf{k})+\left[\epsilon_{ifl}e^{f}_{3}(\textbf{k})e^{l}_{2}(\textbf{k})\right]\otimes e^{j}_{1}(\textbf{k})\Big\}  \\
&&\times\Big\{\left[\epsilon_{apq}e^{p}_{3}(\textbf{k})e^{q}_{1}(\textbf{k})\right]\otimes e^{b}_{2}(\textbf{k})+\left[\epsilon_{apq}e^{p}_{3}(\textbf{k})e^{q}_{2}(\textbf{k})\right]\otimes e^{b}_{1}(\textbf{k})\Big\}.
\eea
According to the cross product relations shown in Eqs. (\ref{cro_ pro_1}) and (\ref{cro_ pro_2}), the  equation  above can be rewritten as
\bea \label{sum_polar_end}
\nonumber
\mathcal{G}_{ijab}(\textbf{k})&=&\left[e^{i}_{2}(\textbf{k})\otimes e^{j}_{1}(\textbf{k})+e^{i}_{1}(\textbf{k})\otimes e^{j}_{2}(\textbf{k})\right]\left[e^{a}_{2}(\textbf{k})\otimes e^{b}_{1}(\textbf{k})+e^{a}_{1}(\textbf{k})\otimes e^{b}_{2}(\textbf{k})\right]   \\
\nonumber&&+\left[e^{i}_{2}(\textbf{k})\otimes e^{j}_{2}(\textbf{k})-e^{i}_{1}(\textbf{k})\otimes e^{j}_{1}(\textbf{k})\right]\left[e^{a}_{2}(\textbf{k})\otimes e^{b}_{2}(\textbf{k})-e^{a}_{1}(\textbf{k})\otimes e^{b}_{1}(\textbf{k})\right]    \\
&=&\sum_{\lambda}e_{ij}(\textbf{k},\lambda)e_{ab}(\textbf{k},\lambda),     
\eea
which can be further calculated by using Eqs. (\ref{orth_ rela}) and (\ref{orth_ 2}) as
\bea \label{sum_polar_end_2}
\nonumber
\mathcal{G}_{ijab}(\textbf{k})&=&\delta_{ia}\delta_{jb}+\delta_{ib}\delta_{ja}-\delta_{ij}\delta_{ab}+\hat{k}_{i}\hat{k}_{j}\hat{k}_{a}\hat{k}_{b}+\hat{k}_{i}\hat{k}_{j}\delta_{ab}+\hat{k}_{a}\hat{k}_{b}\delta_{ij}-\hat{k}_{i}\hat{k}_{a}\delta_{jb}-\hat{k}_{i}\hat{k}_{b}\delta_{ja}  \\  
&&-\hat{k}_{j}\hat{k}_{a}\delta_{ib}-\hat{k}_{j}\hat{k}_{b}\delta_{ia}.
\eea

\end{document}